\documentclass[pra,showpacs,preprintnumbers,amsmath,amssymb,]{revtex4}
\usepackage{bm}
\usepackage{graphicx}
\begin{document}
\title{Spin turbulence in spinor Bose-Einstein condensates}

\author{Makoto Tsubota}
\email{tsubota@sci.osaka-cu.ac.jp}
\author{Kazuya Fujimoto}
\affiliation{Department of Physics, Osaka City University, Osaka 558-8585,  Japan}

\begin{abstract}
We summarize the recent theoretical and numerical works on spin turbulence (ST) in spin-1 spinor Bose-Einstein condensates.
When the system is excited from the ground state, it goes through hydrodynamic instability to ST in which the spin density vector has various disordered direction.  
The properties of ST depend on whether the spin-dependent interaction is ferromagnetic or antiferromagnetic.
ST has some characteristics different from other kinds of turbulence in quantum fluids.
Firstly, the spectrum of the spin-dependent interaction energy exhibits the characteristic power law different from the usual Kolmogorov -5/3 law.
Secondly, ST can show the spin-glass-like behavior; the spin density vectors are spatially random but temporally frozen.
\end{abstract}


\maketitle

\section{Introduction}
Quantum turbulence (QT) is currently one of the most important topics in low temperature physics \cite{PLTP}.
QT has been long studied for superfluid helium, while the realization of Bose--Einstein condensation in trapped atomic gases has proposed another important stage of QT.  
After the realization in 1995 lots of experimental and theoretical works have been devoted to the issues of quantized vortices in this system \cite{Fetterrev2,KasamatsuPLTP, FetterJLTP,AndersonJLTP}.
Most works have addressed the system of a small number of vortices or a vortex array under rotation, but there are very few on QT.
Following a few theoretical proposals \cite{Berloff2002,Parker,Kobayashi2007},  Henn {\it et al.} succeeded in creating and observing three-dimensional QT in trapped $^{87}$Rb BECs \cite{Henn2009a,Henn2009b}.
An important advantage of atomic gases is that multicomponent BECs can be created experimentally. 
Multicomponent BECs allow the formation of various unconventional topological defects 
with complex properties that arise from the internal degrees of freedom and the interactions between different components \cite{Kasamatsurev,Kawaguchireview,Kurnreview}, yielding  a rich variety of superfluid dynamics \cite{Novelsuperfluids,PR}. 
Hence multicomponent BECs can propose a more advanced system of QT.
Two-component  QT has already been studied theoretically and numerically \cite{TakeuchiCSI,Ishino,Karl13}; the most important interest would be how two kinds of QT interact. 

Recently we have studied theoretically and numerically spin turbulence (ST) in spinor BECs \cite{FT12a,FT12b,AT13,TAF13,FT13, AFT13}. 
A spinor BEC is a kind of multicomponent BECs with spin degrees of freedom \cite{Kawaguchireview,Kurnreview}. 
In contrast to the two-component BECs, interatomic interactions allow 
for a coherent transfer of population 
between different hyperfine spin states (spin-exchange collisions), which yields a fascinating physics different from two-component BECs.
Spinor BECs can be another important stage of turbulence in quantum fluids.
When the system is highly excited from the ground state, it goes through hydrodynamic instability to ST in which the spin density vector has various disordered direction.  
This ST shows characteristic behaviors different from other kinds of QT.
First, the spectrum of the spin-dependent interaction energy obeys a -7/3 power law \cite{FT12a,FT12b}, which is different from the traditional Kolmogorov -5/3 power law in turbulence \cite{Frisch,Davidson}.  
Secondly, the spin density vectors are spatially random but temporally frozen, which reminds us of the analogy of spin glass \cite{TAF13}.
This article reviews such recent works on ST.

\section{The Gross-Pitaevskii equations and the hydrodynamic description}
We consider a two-dimensional spin-1 spinor BEC at zero temperature.
 The macroscopic wave functions $\Psi _m$ with the magnetic quantum number $m$ ($m = 1,0,-1$) obey the Gross-Pitaevskii (GP) equations \cite{Ohmi98,Ho98}
 \begin{eqnarray}
 i\hbar \frac{\partial}{\partial t} \Psi _{m} &=& (-\frac{\hbar ^2 }{2M} \nabla ^2 + V) \Psi _{m} 
 +  \sum _{n=-1} ^{1} [- g \mu _{B}(\mathbf{B} \cdot \hat{\mathbf{S}})_{mn} + q(\mathbf{B} \cdot \hat{\mathbf{S}})^{2}_{mn} ]\Psi _{n}  \nonumber \\
 &+& c_{0} \rho \Psi _{m} +  c_{1} \sum _{n=-1} ^{1} \mathbf{s} \cdot \hat{\mathbf{S}} _{mn} \Psi _{n}.  \label{2dGP}
 \end{eqnarray}
Here, $V$ and $\mathbf{B}$ are the trapping potential and magnetic field. The parameters $M$, $g$, $\mu _{B}$, and $q$ are the mass of a particle,
the Land$\rm \acute{e}$ $g$ factor, the Bohr magneton, and a coefficient of the quadratic Zeeman effect, respectively.
 The total density $\rho$ and the spin density vector $s_{i}$ ($i = x, y, z$ ) are given by $\rho =  \sum _{m=-1} ^{1}|\Psi _m|^2$ and  $s_{i} = \sum _{m,n = -1}^{1} \Psi _{m}^{*} (\hat{S}_{i})_{mn} \Psi _{n}$ with
 the spin-1 matrices $(\hat{S}_{i})_{mn}$. 
 The averaged density is given by $\rho_0=\int \rho d\mathbf{r}/A$ with the system area $A$.
 The parameters $c_{0}$ and $c_{1}$ are the coefficients of the spin-independent and spin-dependent interactions.
We focus on the spin-dependent interaction energy ${\cal E}_{s} = \frac{c_{1}}{2} \int \mathbf{s} ^{2} d \mathbf{r}$, whose coefficient $c_{1}$ determines whether the system is ferromagnetic ($c_{1} < 0$) or antiferromagnetic ($c_{1} > 0$).

The GP equations (\ref{2dGP}) allows us to obtain the hydrodynamic equations of the spin density vector, which we need to understand the scaling power law in the energy spectrum.  
The full hydrodynamic equations are derived by Yukawa and Ueda \cite{YU12}, while its simper version is available for the ferromagnetic case \cite{Lamacraft08,Kudo10,Kudo11}.  
We will assume two things.
First, the total density $\rho$ is time-independent because $c_0$ is usually much larger than $c_1$.
Secondly,  the spin density vector takes the largest value everywhere owing to the ferromagnetic interaction.
The two assumptions are confirmed in our numerical simulation \cite{FT12a}.  
Then the normalized spin density vector $\hat{\mathbf{s}} = \mathbf{s} /\rho_{0}$  is shown to obey the hydrodynamic equations 
\begin{equation}
\frac{\partial}{\partial t} \hat{\mathbf{s}} + (\mathbf{v} \cdot \mathbf{\nabla}) \hat{\mathbf{s}} = \frac{\hbar}{2M} \hat{\mathbf{s}} \times [ \nabla ^{2} \hat{\mathbf{s}} + (\mathbf{a} \cdot \mathbf{\nabla}) \hat{\mathbf{s}} ], \quad \mathbf{v} = \frac{\hbar}{2iM\rho} \sum _{m=-1} ^{1} ( \psi _{m}^{*} \mathbf{\nabla} \psi _{m} - \psi _{m} \mathbf{\nabla} \psi _{m}^{*} ),
\end{equation}
where $\mathbf{a} = (\nabla \rho)/\rho$ and $ \mathbf{v}$ is the superfluid velocity \cite{Lamacraft08,Kudo10, Kudo11}. 
In a uniform system $\rho$ keeps almost constant, thus vanishing $\mathbf{a}$.
As a result, we obtain
\begin{equation}
\frac{\partial}{\partial t^{'}} \hat{\mathbf{s}} + \Bigl(\frac{\mathbf{v}}{c_{s}} \cdot \mathbf{\nabla} ^{'}\Bigr) \hat{\mathbf{s}} =  \hat{\mathbf{s}} \times \mathbf{\nabla} ^{' 2} \hat{\mathbf{s}}.  \label{s}
\end{equation}
Here space and time are respectively normalized by the coherence length $\xi = \hbar /\sqrt{2Mc_{0}\rho_0}$ and a characteristic time $\tau = \hbar /c_{0}\rho_0$ ($t^{'} = t/\tau$, $\mathbf{\nabla}^{'} =  \xi \mathbf{\nabla}$), and $c_{s} = \sqrt{c_{0}\rho_{0}/2M}$ is the sound velocity. 
Another important characteristic length is the spin coherence length $\xi _{s} = \hbar / \sqrt{2M|c_{1}| \rho_{0}}$, which is relevant to spin structures such as domain walls and polar core vortex \cite{Saito07}.

\section{Spin turbulence with the ferromagnetic interaction}
We numerically created ST by three different methods: (1) the instability of counterflow between the $m=\pm 1$ components in a uniform system \cite{FT12a}, (2) the instability of the helical spin structure in a trapped system \cite{FT12b}, (3) the instability by an oscillating magnetic field in a uniform system \cite{AT13}. 
All simulations are performed in a two-dimensional system.
In every case we confirmed the -7/3 power law in the spectrum of the spin-dependent interaction energy and the spin-glass-like behavior \cite{TAF13}, though only in the case (3) how to apply the oscillating magnetic field is different between \cite{AT13} and \cite{TAF13}.
In this section we describe the case of (1) as a typical one.

\subsection{Spin turbulence and the energy spectrum}
The counterflow between the $m = \pm  1$ components with a relative velocity $\mathbf{V}_{R} = V_{R} \hat{\mathbf{e}}_{x}$ is represented by the initial wave functions
\begin{equation}
(\Psi _1, \Psi _0, \Psi _{-1})
= \sqrt{\frac{\rho_{0}}{2}} \Bigl({ \rm{exp}}\Bigl [i\Bigl(\frac{M}{2 \hbar}\mathbf{V}_{R}\cdot \mathbf{r} - \frac{\mu _{1}}{\hbar}t\Bigr)\Bigr],  0,  {\rm{exp}} \Bigl[-i\Bigl(\frac{M}{2 \hbar}\mathbf{V}_{R}\cdot \mathbf{r} + \frac{\mu _{-1}}{\hbar}t\Bigr)\Bigr]\Bigr), 
\end{equation}
where $\hat{\mathbf{e}}_{x}$ is a unit vector along the $x$ direction and the chemical potentials $\mu _{1}$ and $\mu _{-1}$ are equal to $c_{0}\rho_{0} + MV_{R}^2/8$.
The analysis of the Bogoliubov-de Gennes equation shows that the system is dynamically unstable at any relative velocity $V_R$ \cite{FT12a}.
Hence the initial counterflow goes through the dynamical instability towards ST.

\begin{figure}[h]
\includegraphics[width=38pc]{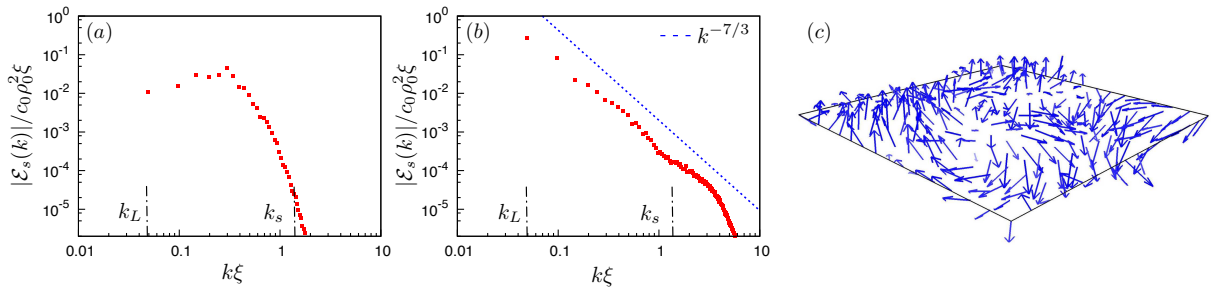}
\caption{\label{Ferro}Time dependence of spectrum of spin-dependent interaction energy ${\cal E}_{s}(k)$ for the ferromagnetic case at $t/\tau =$ (a) $150$, and (b) $3500$. The dotted line in (b) refers to the $k^{-7/3}$ power law. The wave numbers $k_L$ and $k_s$ correspond to the system size and the spin coherence length, respectively. (c):Profile of the spin density vector corresponding to (b).
This ST is obtained through the counterflow instability with $c_{0}/c_{1} = -20$, $c_{0} > 0$, $V_{R}/c_{s} = 0.78$.
The coordinate is normalized by the coherence length $\xi$ and the box size is $128\times 128$. 
Space in the $x$ and $y$ directions is discretized into $512 \times 512$ bins. }
\end{figure}

How ST develops is shown in Fig. \ref{Ferro}.  
The energy spectrum ${\cal E}_s(k)$ has in the early stage a peak corresponding to the most unstable mode of the dynamical instability (Fig. \ref{Ferro} (a)), while the energy gradually flows to the higher wave numbers. 
When the spin density vector is enough disordered as shown in Fig. \ref{Ferro} (c), the energy spectrum takes a characteristic -7/3 power law (Fig. \ref{Ferro} (b)). 

Such a characteristic power law exhibits that this ST is not just a random disordered state but an organized self-similar state. 
We can find easily the similar scenario in classical turbulence (CT) \cite{Frisch,Davidson}.  
Three-dimensional fully developed turbulence has the inertial range, where the kinetic energy is transfered selfsimilarly from low to high wave numbers and its spectrum shows the Kolmogorov -5/3 law. 
This Kolmogorov law can be obtained by applying the scaling analysis due to the selfsimilarity to the hydrodynamic equations \cite{scaling2}.
The similar derivation is available to the hydrodynamic equation (\ref{s}).
The present case allows us to neglect the second term in the left hand side, because $|\mathbf{v}|$ is much smaller than $c_s$.  
Let us assume that Eq. (\ref{s}) is invariant for the scale transformation  $\mathbf{r}\rightarrow \alpha \mathbf{r}$ and $ t \rightarrow \beta t$. Then it is easy to find that the spin density vector should be transformed as  $\hat{\mathbf{s}} \rightarrow \alpha^2 \beta^{-1} \hat{\mathbf{s}} $.
Hence we obtain $\hat{\mathbf{s}} \sim \Lambda k^{-2} t^{-1}$ with a dimensional constant $\Lambda$ in the scaling region.
We apply the usual dimensional analysis to the scaling region to suppose that the energy flux $\epsilon \sim \hat{\mathbf{s}}^2 t^{-1} \sim \Lambda^2 k^{-4} t^{-3}$ is independent of $k$.
Then the energy spectrum ${\cal E}_s(k)$ should be represented only by $\epsilon$, $k$ and $\Lambda$, resulting in  
the spectrum ${\cal E}_s(k) \sim  \hat{\mathbf{s}}^2 k^{-1}  \sim \Lambda^{2/3} \epsilon ^{2/3} k^{-7/3}$.

\subsection{Spin-glass-like behavior}
ST shows another characteristic behavior. 
In ST the spin density vectors become spatially random but temporally frozen, which reminds us of spin glass.
Spin glasses are magnetic systems in which the interactions between the magnetic moments are in conflict with each other \cite{RMP}.
Thus, these systems have no long-range order but exhibit a freezing transition to a state with a kind of order in which the spins are aligned in random directions.

\begin{figure}[h]
\includegraphics[width=18pc]{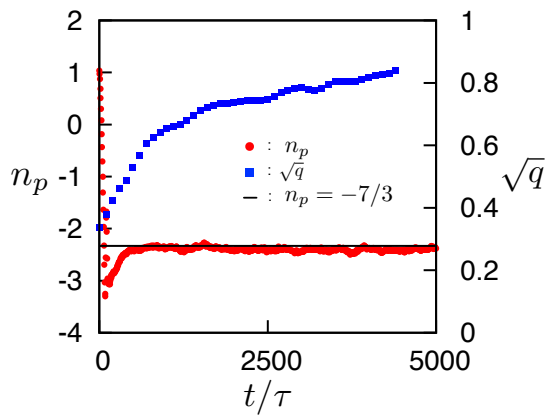}\hspace{2pc}%
\begin{minipage}[b]{14pc}\caption{\label{ST}Time development of the exponent $n_{p}$ of the spectrum ${\cal E}_s(k)$ and the spin glass order parameter $ \sqrt{q(t)} $.  The conditions and the parameters are same as those of Fig. \ref{Ferro}. See  Ref. \cite{TAF13} for the detail.}
\end{minipage}
\end{figure}

In order to characterize the spin-glass-like behavior in ST, we introduced an order parameter following the spin-glass theory \cite{RMP,SK}.
We should note that the order parameter is time-dependent.
For the normalized spin density vector  $\hat{\mathbf{s}}(\mathbf{r},t)=\mathbf{s}(\mathbf{r},t)/|\mathbf{s}(\mathbf{r},t)|$, its space average and the time average during the period $[t, t+T]$ could be defined as 
\begin{equation}
 [\hat{\mathbf{s}} (\mathbf{r},t)]=\frac1A \int_A \hat{\mathbf{s}} (\mathbf{r},t)d\mathbf{r}, \quad  \langle \hat{\mathbf{s}} (\mathbf{r},t)\rangle_T= \frac1T \int_t^{t+T}\hat{ \mathbf{s}} (\mathbf{r},t_1)dt_1.
 \label{average}\end{equation}
The argument of how to take $T$ is discussed in Ref. \cite{TAF13}.
The time-dependent order parameter is defined as 
\begin{equation}
q(t)=[\langle\hat{ \mathbf{s}} (\mathbf{r},t)\rangle_T^2]. \label{q}
\end{equation}
If the spin density vector is completely frozen, $q(t)$ should be unity. Figure \ref{ST} shows the time dependence of $q(t)$ and the power exponent $n_p(t)$ of the spectrum of the spin-dependent interaction energy. 
The order parameter $q(t)$ increases obviously as the exponent $n_{p}(t)$ approaches $-7/3$, which means that the spin-glass-like order grows as the ST with the $-7/3$ power law develops. 
Of course, our system of spinor BECs differs from a magnetic system yielding spin glass. 
Spin glass states do not show such a power law behavior in the energy spectrum, being much different from ST. 
We do not know currently what causes the spin-glass-like behavior in ST \cite{TAF13}.

\section{Spin turbulence with the small spin magnitude}
When the amplitude of spin density vector is small, another -1 power law appears in the low wave number region in addition to the -7/3 power in the high wave number region of the  energy spectrum in ST \cite{FT13}.
 This kind of ST is realized in  two cases:
(i)  with antiferromagnetic  interaction and (ii) with ferromagnetic  interaction under a static magnetic field.
In the first article \cite{FT12a} we found that the antiferromagnetic interaction allowed the system to develop to ST too but  the power exponent of ${\cal E}_s(k)$ did not obey the simple scaling behavior discussed in the last section. 
Investigating the behavior of  the ST led us to the consideration that the different behavior was attributable to the small spin magnitude.
Then we noticed that ST with small spin amplitude is realized in the system with ferromagnetic  interaction under a static magnetic field too, where the quadratic Zeeman term reduces the spin amplitude.

The scaling argument should start with the full hydrodynamic equation \cite{YU12} described by $\mathbf{s} $ and the nematic tensor
\begin{equation}
n_{\mu \nu} =  \frac{1}{\rho} \Psi _{m}^{*} (\hat{N}_{\mu \nu})_{mn} \Psi _{n} \label{nematic}
\end{equation}
with 
\begin{eqnarray}
(\hat{N}_{\mu \nu})_{mn}  = \frac{1}{2} [ (\hat{S}_{\mu})_{ml} (\hat{S}_{\nu})_{ln} + (\hat{S}_{\nu})_{ml} (\hat{S}_{\mu})_{ln} ]. 
\end{eqnarray}
The detail is described in Ref. \cite{FT13}. Here we summarize the results.

\begin{figure}[h]
\includegraphics[width=38pc]{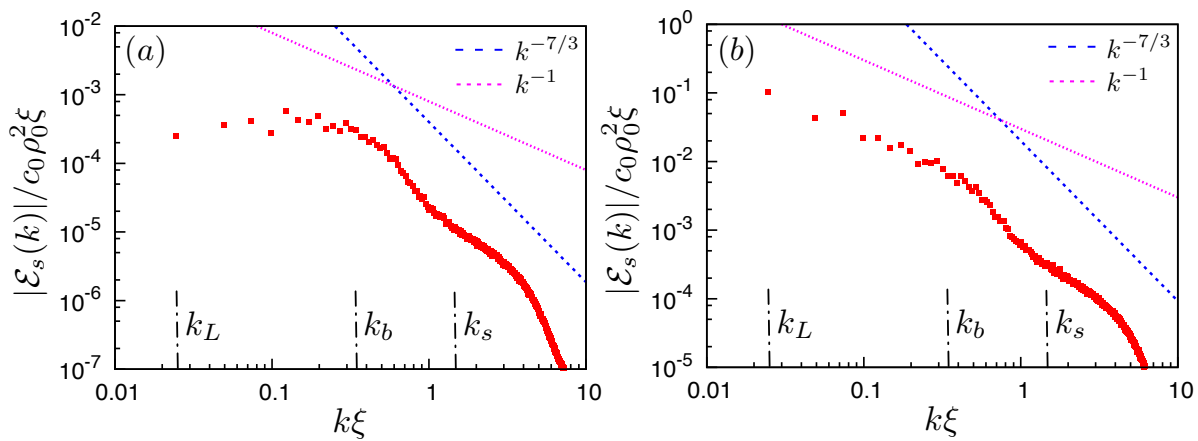}
\caption{\label{small}Spectrum of spin-dependent interaction energy ${\cal E}_s(k)$ for (a) antiferromagnetic interaction at $t/\tau=3500$ and (b) ferromagnetic interaction under a static magnetic field at $t/\tau=3000$ \cite{FT13}. The wave numbers $k_L$ and $k_s$ correspond to the system size and the spin coherence length, respectively, and the boundary wave number  $k_b$ is given in the text.[Fujimoto {\it et al.}: Phys. Rev. A {\bf 88} (2013), reproduced with permission.
Copyright 2013 by the American Physical Society].}
\end{figure}

We apply three approximations. 
The first is that the total density $\rho$ is uniform. 
The second is that the superfluid velocity is much smaller than the sound velocity.
The third is that the magnitude of the spin vector $\mathbf{s}/\rho$ is smaller than unity.
Then appears a boundary wave number $k_{b} = 2 \sqrt{|c_{1}| M \rho_0}/\hbar$. 
Applying the scaling argument to the hydrodynamic equations leads to the spectrum ${\cal E}_s(k) \sim \epsilon_L^{2/3}k^{-1} $ for $k<k_b$ and  ${\cal E}_s(k) \sim \epsilon_H^{2/3}k^{-7/3} $ for $k>k_b$ with different energy flux $\epsilon_L$ and $\epsilon_H$. 
We numerically obtain ST in a uniform system by the counterflow instability. 
The energy spectrum for developed ST is shown in Fig. \ref{small}.
In the system with antiferromagnetic interaction (Fig. \ref{small} (a)),  the spectrum tends to exhibit the expected $-7/3$ power law in the high-wave-number region $k_{b}<k<k_{s}$, but it deviates from the $-1$ power in the low-wave-number region $k_{L}<k<k_{b}$.
The system with ferromagnetic interaction under a static magnetic field exhibits more clearly both the -1 and -7/3 power laws as shown in Fig. \ref{small} (b).  
 
We investigated the order parameter $q$ of Eq. (\ref{q}) for ST with antiferromagnetic interaction too \cite{AFT13}. 
The order parameter $q$ does not grow, which means that spin density vectors are not frozen temporally but fluctuate.
This behavior is much different from the ferromagnetic case.

\section{Conclusions and discussions}
Spinor Bose-Einstein condensates propose a novel system of turbulence in quantum fluids, namely spin turbulence (ST) in which the spin density vectors are disordered.
When the dynamics starts from an initial state with the energy much higher than the ground state or the system is excited from the ground state by an applied field, ST appears after hydrodynamic instability.
ST exhibits some characteristic statistical laws, and the behavior depends on the spin-dependent interaction.
When the spin-dependent interaction is ferromagnetic,  the spectrum of the interaction energy shows the -7/3 power law; this is understood through the scaling analysis for the hydrodynamic equations of spins and confirmed by the numerical simulation of the GP equations.
Furthermore, the spin density vectors are spatially random but temporally frozen, which reminds us of the analogy with spin glass; the order parameter certainly grows in ST.
However, the story is just more complicated when the interaction is antiferromagnetic.
Then appears a boundary wave number $k_b$, and the scaling analysis in ST finds that the energy spectrum obeys the -1 power law for $k<k_b$ and the -7/3 law for $k>k_b$; the -7/3 power law was confirmed numerically.
This energy spectrum actually comes from the small spin amplitude due to the antiferromagnetic interaction. 
The similar situation is expected to occur also for the ferromagnetic system under a static magnetic field that reduces the spin amplitude, in which both
-1 and -7/3 power laws were confirmed clearly by the numerical simulation.
The spin-glass-like behavior did not appear in the antiferromagnetic ST.

Such a spin-disordered state was experimentally created in a trapped system through the instability of the initial helical structures of spin, and actually observed \cite{Kurn08}.
Creating ST and observing the spin density vector $ \mathbf{s}( \mathbf{r},t)$ would allow us to investigate the energy spectrum ${\cal E}_s(k)$ and the spin-glass-like behavior. 
  
It is very important to ask what causes the energy cascade in ST.
A mechanism causing the Kolmogorov $-5/3$ power law in classical turbulences is considered to be the Richardson cascade of vortices \cite{Davidson, Frisch}.
Since there are very few topological defects in our simulation of ST, however,  the cascade due to them is not relevant. 
At present we guess the cascade of spin waves plays an important role for ST.
This will be worked in reference to the wave turbulence theory \cite{wt1, wt2}.

Usual selfsimilar power law in the inertial range of turbulence is caused by some dissipation at high wave numbers, which makes the energy flow towards high wave numbers. 
Our system based on Eq. (\ref{2dGP}) conserves the total energy; no dissipative mechanisms are included in our formulation.
Thus it is not so trivial to understand why and how the characteristic power laws are built and sustained in ${\cal E}_s(k)$. 
Here we should note that spinor BECs have superfluid degrees of freedom (mass current) as well as spin degrees of freedom and they are coupled.
Hence the spin-dependent interaction energy ${\cal E}_s$ is not conserved in the dynamics.
While spin develops ST, superfluid may create its own turbulence. 
It would be quite interesting to investigate how these degrees of freedom interact with each other and develop turbulence for the whole system.

\section*{ACKNOWLEDGMENT}
The authors thank Yusuke Aoki for the scientific collaboration and the preparation of the manuscript.

\end{document}